\RequirePackage{fix-cm}
\documentclass[smallextended]{svjour3}       % onecolumn (second format)
\smartqed  % flush right qed marks, e.g. at end of proof
\usepackage{graphicx}
\usepackage{color,verbatim,enumerate,amsmath,amssymb}
 \usepackage{mathptmx,float}
 \usepackage{natbib}
\setcitestyle{authoryear,open={(},close={)}}
\begin{document}
\newcommand{\TT}{\mathbf{T}}

\title{A nonparametric graphical tests of significance in functional GLM}
\thanks{The project has been financially supported by the Grant Agency of Czech Republic (Project No. 19-04412S).}


\author{Tom\'a\v{s} Mrkvi\v{c}ka       \and
Tom\'a\v{s} Roskovec
\and
Michael Rost
}


\institute{All authors \at
              Dpt. of Applied Mathematics and Informatics, Faculty of Economics, University of South \\ Bohemia, Studentsk{\'a} 13, 37005 {C}esk\'e Bud{e}jovice, Czech Republic \\
              Tel.: +420-387772700\\
              \email{mrkvicka.toma@gmail.com}           %  \\
%             \emph{Present address:} of F. Author  %  if neede
}

\date{Received: date / Accepted: date}

\maketitle

\begin{abstract}
A new nonparametric graphical test of significance of a covariate in functional GLM is proposed. Our approach is especially interesting due to its functional graphical interpretation of the results. As such it is able to find not only if the factor of interest is significant but also which functional domain is responsible for the potential rejection. In the case of functional multi-way main effect ANOVA or functional main effect ANCOVA models it is able to find which groups differ (and where they differ), in the case of functional factorial ANOVA or functional factorial ANCOVA models it is able to find which combination of levels (which interactions) differ (and where they differ).  The described tests are extensions of global envelope tests in the GLM models. It applies Freedman-Lane algorithm for the permutation of functions and as such it approximately achieve the desired significance level. 
\end{abstract}

% % \section{Introduction}
% \label{intro}
% Your text comes here. Separate text sections with
% \section{Section title}
% \label{sec:1}
% Text with citations \citep{RefB} and \citep{RefJ}.
% \subsection{Subsection title}
% \label{sec:2}
% as required. Don't forget to give each section
% and subsection a unique label (see Sect.~\ref{sec:1}).
% \paragraph{Paragraph headings} Use paragraph headings as needed.
% \begin{equation}
% a^2+b^2=c^2
% \end{equation}

%\begin{figure}
%  \includegraphics{example.eps}
%\end{figure}
%\begin{figure*}
%  \includegraphics[width=0.75\textwidth]{example.eps}
%\end{figure*}
%\begin{table}
%\begin{tabular}{lll}
%\hline\noalign{\smallskip}
%\end{tabular}
%\end{table}

\section{Introduction}
Functional general linear models (GLM) appear in various scientific fields, where the observed data are in the form of function, e.g. in medicine, finance, biology, etc. In this paper, we will consider a $d$-dimensional functions and study their dependence on various factors - continuous, categorical and also interactions, through a GLM. 

The problem of functional GLM is widely studied in the literature. 
For example \citep{RamsaySilverman2006} described a bootstrap procedure based on pointwise $F$-tests, \citep{abramovich2006testing} used wavelet smoothing techniques, and \citep{FerratyEtal2007} used dimension reduction approach. Further, \citep{Cuesta-Albertos2010} applied the $F$-test on several random univariate projections and bound the tests together through false discovery rate.

Since we deal with functional data the assumptions for parametric methods are more complex and in practice not guaranteed to be fulfilled, therefore the nonparametric methods, which have much fewer assumptions, are very popular in this area, e.g. \citep{NicholsHolmes2001}, \citep{Winkler2014}, \citep{PantazisEtal2005} concentrate on certain pointwise statistics, such as the $F$-statistic, and find the distribution of its maxima by permutation. \citep{Hahn2012} used a univariate integral deviation statistic to summarise the deviances between groups in one-way ANOVA. All these permutation methods find a pure maximum of a statistic or compute the integral from a statistic over the study domain. It requires that the statistic has homogeneous distribution across the functional domain which is not necessarily the case in the practice. 
On the other hand, our nonparametric procedure solves this problem by applying non-parametric rank envelope test \citep{MyllymakiEtal2017}, \citep{MrkvickaEtal2017} instead.

The proposed method is an extension of the method described first in \citep{MrkvickaEtal2019}. The original description considers only one-way functional ANOVA model, whereas here we concentrate on the application of this method for general linear model designs with retaining of the interesting graphical interpretation of the results. This graphical interpretation is able to show which 2 groups differ together with showing in what area of the functions these groups differ without application of any posthoc test in case of a categorical factor of interest. It is able to indicate which area is responsible for the rejection of a continuous factor of interest. It is also able to show which combinations of levels in multi-way ANOVA design or multi-way ANCOVA design are responsible for the rejection of the interaction factor. By this, the interpretation of analysis is much easier. The graphical visualisation of such a generality was not introduced to this field before by our knowledge.

Our nonparametric graphical test is exact, i.e. the true significance level of the test is the pre-set significance level of the test, if only one factor is in the analysis, because the exchangeability of the permutation is fulfilled. When adding the nuisance factors, the exchangeability of the permutations is not straightforward. Certain permutation strategies have to be applied in order to achieve the exact test \citep{Winkler2014}. In our approach we follow the recommendation accepted in the univariate permutation tests \citep{AndersonBraak2003}, \citep{Winkler2014}, i.e. to apply the permutation of residuals under the reduced model which was first described in \citep{FreedmanLane1983}. This permutation strategy is not exact but comes the closest to the conceptually exact level \citep{AndersonRobinson2001} and performed the best in various circumstances \citep{AndersonLegendre1999}.

The organisation of the paper is as follows. In Section \ref{sec:Th} we specify the mathematical setting for our method and how every step of algorithm works. First, we discuss the exchangeability property and usage of Freedman--Lane algorithm. Then we describe the Global extreme rank length envelope test and how it is applied on test vectors. 

In Section \ref{sec:SS}, we perform a simulation study to compare the powers of our graphical procedures with the powers of the procedures which are already available. We have chosen the random projection method (RPM) which is available through the software R \citep{Rfda.usc} and $F$-max method \citep{NicholsHolmes2001}. Also, we show the graphical output for the categorical predictor, continuous predictor and interactions. 
%All our proposed methods can be found in the R package GET, which is available at github (\texttt{https://github.com/myllym/GET}).
The Section \ref{sec:DC} is left for conlusions and discussion.

\section{Theory}\label{sec:Th}
\subsection{Mathematical specification}
We study a set of functions (possibly multidimensional) ${\bf Y}=(y_1(t), \ldots , y_n(t))^t$, $t\in D\subset \mathbb{R}^d$. We consider several exploratory factors of these functions and set a general linear model by two matrices ${\bf X}=(x_{ij}(t))_{i=1,\ldots , n, j=1,\ldots , J}$ (defining the factor of interest) and ${\bf Z}=(z_{ij}(t))_{i=1,\ldots , n, j=1,\ldots , Z}$ (defining nuisance factors).  Generally, we assume that the factors can be functional, but usually the factors are constants (e.g. age, gender, ... ). 

By {\it full model} we understand
$${\bf Y}={\bf X}\beta+{\bf Z}\gamma+\varepsilon,$$
where $\beta$ and $\gamma$ are parameters of GLM calculated in every $t \in D$. Thus $\beta=(\beta_{j}(t))_{j=1,\ldots , J}$ and $\gamma=(\gamma_{j}(t))_{j=1,\ldots , Z}$ are $J$ and $Z$ dimensional functions representing an effect of factors. $\varepsilon=(\varepsilon_{1}(t), \ldots , \varepsilon_{n}(t))^t$ is the set of residuals.

By {\it null model} we understand 
$${\bf Y}={\bf Z}\gamma+\varepsilon,$$
i.e. we ignore ${\bf X}$ and regress considering only the nuisance factors. 

The elements of $\bf{Y}$ represent functions $y(t)$ depending on variable $t$, possibly more than one dimensional. But since the data have to be discretised anyway, we consider only finitely many values of $t$ and therefore $y\in\bf{Y}$ is a vector of values of the function in question. For simplicity, we refer to the variable $t$ as ``time'' further in this paper, regardless of specific data meaning and the dimension of this variable.

We assume that the discretization is done for same $K$ values of $t$ for every function $y(t)$. That means $\bf{Y}$ is a multi-vector of size $n\times K$, $\beta$ is a multi-vector of size $J\times K$, $\gamma(t)$ of size $Z\times K$ and $\varepsilon$ of size $n\times K$. 

Our study aims to test null hypothesis of the significance of the factor contained in the matrix $\bf{X}$, i.e.
$$H_0: \beta_{jk} = 0, \text{ for all } j=1,\ldots , J, k=1,\ldots , K.$$
Remark that both $\bf{X}$ and $\bf{Z}$ may include all kinds of factors, both continuous and categorical ones and even the interactions of effects.

\subsection{Choice of methods}
Our goal is to test $H_0$ by non-parametric permutation method with graphical interpretation. We based it on two advanced results. First, the permutation scheme with the presence of nuisance effects is based on  {\it Freedman-Lane algorithm} \citep{FreedmanLane1983}. Second, we visualise the relationship between the original and the permuted data by the {\it global extreme rank length envelope test}  \citep{MrkvickaEtal2017,MyllymakiEtal2017}.

\subsection{ANOVA effect study}

Let us begin with univariate one way ANOVA. Then the data consist of several groups of samples, and we decide if there is a difference between the groups or if the sorting has no effect. 
The general nonparametric permutation method suggests permuting the samples between groups randomly. Then we compare the initial sorting to the sorting of new permuted data, for example by $F$-test statistics. The result should be the same if the belonging to a group does not affect sample within. The necessary property of data and permutation algorithm is that the permutations do not change the joint distribution of tests statistic under the null hypothesis. This property is called {\it exchangeability}.

The property of {\it exchangeability} is lost if we add a nuisance effect. For example, if we want to test medicine results of several treatments and the result depends on the age of a patient, the permutation may sort all old patients data into one group and all young patients data into the other. Then we may wrongly conclude the result based on the age and not on the used medicine. There are several ways how to avoid this situation, for example, we may restrict the family of admissible permutations or bound the samples into vectors of samples in order to keep information about nuisance effect despite permuting. We avoid this mistake by the Freedman-Lane algorithm described further, recommended in the univariate GLM permutation studies \citep{AndersonBraak2003}. Suggested Freedman-Lane algorithm permutes the residuals under the null model instead of permuting data itself. The exchangeability property is not satisfied with this scheme, but the significance level for such permutation procedures is close to the nominal one while retaining good power of the tests as it will be shown by the simulation study.

\subsection{Freedman-Lane algorithm}
\begin{enumerate}
\item Regress data ${\bf Y}$ against the full model containing both the effect of interest $\beta$ and the nuisance effect $\gamma$ as
$${\bf Y}={\bf X}\beta+{\bf Z}\gamma+\varepsilon.$$
\item Regress data ${\bf Y}$ against the reduced model containing only the nuisance effect $\gamma$ as
$${\bf Y}={\bf Z}\gamma+\varepsilon_{{\bf Z}}.$$
\item Compute the permuted data ${\bf Y}^*_j$. We get this data by permuting the residuals of reduced model $\varepsilon_{{\bf Z}}$ by permutation $\pi_j$ into $\varepsilon_{{\bf Z},j}=\pi_j(\varepsilon_{{\bf Z}})$. We get
$${\bf Y}^*_j={\bf Z}\gamma+\varepsilon_{{\bf Z},j}.$$
\item Regress the permuted data ${\bf Y}^*_j$ against the full model and get a new effect of interest $\beta^*_j$ from formula
$${\bf Y}^*_j={\bf X}\beta^*_j+{\bf Z}\gamma^*_j+\varepsilon_{j}$$
\end{enumerate}

\subsection{Global extreme rank length envelope test}
The other specification of our work is the graphical interpretation and finding the differing groups without the need of post-hoc test in case of the categorical factor of interest or interactions. If there is a nonzero effect of interest for a categorical factor, there is the need of post-hoc test to find the responsible party. But the post-hoc procedure may slightly differ from the result of the ANOVA test. Our method provides a $(1-\alpha) 100\%$ global envelope,  which contains simultaneously all parameters $\beta_{jk}, j=1, \ldots, J, k=1, \ldots, K.$ So if the test vector is completely contained in the envelope for every time, the null hypothesis is not rejected. If the test vector leaves the envelope in any value, the null hypothesis is rejected, and this group is identified as the group responsible for rejecting the null hypothesis. Moreover, since we study functions, we see the values of time for whose this group does not fit into the envelope and may use this information for interpretation of data and conclusions. The positions of the envelopes and the test vectors can be visualised, and the results are easy to be interpreted.

\subsubsection{Test vector choice}\label{DefinujemeTestoveVektory}
To apply the rank envelope test, we have first to select a test vector. 
The first possible choice is, based on the values of effects 
$$\mathbf{T}=(\beta_{jk}), j=1, \ldots , J, k=1, \ldots , K.$$
That is a multi-vector $J\times K$ for $J$ different groups of the factor of interest and $K$ values of time. I.e. for continuous factor of interest $J=1$ and $\mathbf{T}=(\beta_{1k}), k=1, \ldots , K$. For the categorical factor of interest, $J$ is equal to the number of groups of the categorical factor, with each $\beta_j$ having the same role in the univariate model, adding the additional condition of $\sum_j \beta_j=0$. For interaction of continuous and categorical factor, $J$ is also equal to the number of groups of the categorical factor, adding the same additional condition of $\sum_j \beta_j=0$. For the interaction of two categorical factors, $J$ is equal to the product of the numbers of groups of both categorical factors, adding the same additional condition of $\sum_j \beta_j=0$.
To construct the envelope, we consider test vectors of the same type based on permuted data.

The second possible choice of the test vector, applicable for at least one categorical factor,  consists of differences between two group parameters of $\beta$. Since we do not have to check the couples $i=j$ and $i>j$, we check only the couples where the first index is lower than the second one. We get a different test vector
$$\begin{aligned}\mathbf{T}'=(\beta_{11}-\beta_{21}, \ldots, \beta_{1K}-\beta_{2K},& \beta_{11}-\beta_{31}, \ldots \\
\ldots, &\beta_{1K}-\beta_{3K}, \ldots, \beta_{(J-1)1}-\beta_{J1}, \ldots, \beta_{(J-1)K}-\beta_{JK}).
\end{aligned}$$
The number of coordinates of $\mathbf{T}'$ is $(J(J-1)/2)\times K$. 
%Both previous tests are done on the same level of significance $\alpha$ and 
For the first test vector $\mathbf{T}$ we get the information about the groups causing possible rejection of the null hypothesis. For the second test vector $\mathbf{T'}$ we get the information about the couples of groups which are different. Although we ask the same question about the data, either test is sensitive to the different kind of misfits.

In case of unequal variances among groups of functions, in case of categorical factors either nuisance or one of interest or categorical interactions, we may use the group variance normalising transformation before the analysis as described in \citep{MrkvickaEtal2019}. The check for homoscedasticity of the groups in the presence of nuisance factors can be done in the same way as in \citep{MrkvickaEtal2019}, where it is described without nuisance factors.

\subsubsection{Global extreme rank length envelope test}\label{PopisGET}
We briefly introduce the method developed in \citep{MyllymakiEtal2017} and \citep{MrkvickaEtal2017}. We suppose now that the type of test vector has been chosen. Let us suppose we have $I$ permutations produced by Freedman-Lane algorithm. Let us denote $\TT_i$ the test vector based on $i$-th permutation, especially $\TT_1$ is the vector based on identical permutation or the original data. Using the permuted data, we aim to define boundary vectors $\TT_{upp}$ and $\TT_{low}$. These vectors create a natural envelope including the typical values and excluding the extreme values. We do not reject the null hypothesis $H_0$ if for every element $k$ it holds
$$\TT_1(k)\in(\TT_{low}(k),\TT_{upp}(k)).$$
We reject the hypothesis $H_0$, if there exists an element $k$ such that
$$\TT_1(k)\notin(\TT_{low}(k),\TT_{upp}(k)).$$

We proceed to show how to set $T_{upp}$ and $T_{low}$ and a $p$-value of the test.
First, let us calculate for every vector $\TT_i(k)$ the ranks of its values for each element $k$ separately and denote this value $S_{i}(k)$. 
As we use two side envelope, we search for the extremeness from both sides, i.e. $R_{i}(k)=\min(S_{i}(k),n-S_{i}(k)+1)$.
As an {\it extreme rank} we denote
$$R_{i}=\min_{k\in \{1,2,\dots K\}} R_{i}(k).$$
There is a huge risk of ties, so we have to decide which of two vectors with the same {\it extreme rank} is more centred. We break these ties and order the vectors by so-called {\it extreme rank length}.

First, we order the ranks for every test vector $T_i(k)$ into nondecreasing sequence of $K$ numbers $\mathbf{R}_i$, starting with {\it extreme rank} $R_{i}$. More precisely we set
$$\mathbf{R}_i=(R_{i}(k_1),R_{i}(k_2)\dots R_{i}(k_K)),\text{ such that }R_{i}(k_l)\leq R_{i}(k_{l+1})\text{ for }l\in\{1,2\dots K-1\}.$$ 
Then we define the {\it extreme rank length} relation $\prec$ by: $$\TT_i\prec \TT_{i'},$$ if there exists $n>0$ such that the ranks $R_{i}(k_l) = R_{i'}(k_l)$  for the first $n-1$ elements and $R_{i}(k_n) < R_{i'}(k_n)$.
Roughly speaking we compare the most extreme rank of vectors, and in case of a tie, we compare how many values are so extreme for said vectors. If there is another tie in the length of the most extreme rank, we compare the second most extreme rank, then we compare its length and so on until there is the difference.

Due to the above ordering, we may set the $p$-value as $$p=1-(I^{-1}\sum_{i=1}^I\mathbf{1}(\TT_1\prec \TT_{i})).$$ 

Now we are ready to construct the $(1-\alpha)$ envelope, where $\alpha$ is the preset significance level of the test of the null hypothesis. Remark here that due to the nuisance factors and the Freedman and Lane permutation strategy the probability that $\TT_1(k)$ leaves this envelope is only approximately equal to $\alpha$.    
Let $\mathcal{I}_{ex}\subset \{1,\dots I\}$, $|\mathcal{I}_{ex}|=I\alpha$ denotes the set indexes which corresponds to $I\alpha$  the most extreme vectors. Then we define the {\it global extreme rank length envelope} for all $k$ as 
$$(T_{low}(k),T_{upp}(k))=(\textup{min}_{i\in\{1,\dots I\}\setminus \mathcal{I}_{ex}}\{T_i(k)\},\textup{max}_{i\in\{1,\dots I\}\setminus \mathcal{I}_{ex}}\{T_i(k)\}, \quad k\in 1,\ldots, K.$$

\section{Examples with simulation study} \label{sec:SS}

This Section aims to show several different functional GLM where our methods are applicable and to compare the powers with other non-graphical methods available for functional GLM. Namely with random projection method (RPM) of \citep{Cuesta-Albertos2010} and with $F$-max test \citep{NicholsHolmes2001}, which is often used in neuroimage data analysis.

For this purpose we use a model function which consists of three summands. By the use of this model function, we will simulate all our examples. 
$$\begin{aligned}
y_{i,j,k}(t)&=3(5+2i)t(1-t)^{5+2i}+(\text{max}\{0;64(1-t)(t-0.75)\})^j+t(1-t)k/100\\
&=y_i(t)+y_j(t)+y_k(t).
\end{aligned}$$
We describe the design of our model function now. There are three parameters $i,j,k$ that may represent the effects of three different factors. In every of our example we define the data function by picking a triplet $i,j,k$ and adding the error term $e(t)$ to function $y_{i,j,k}(t)$. The domain of functions is set to $t\in [0,1]$ and it is discretized into 100 equidistant values for every function. If the parameter represent categorical factor, the choice of value for parameter represents an involvement of data function to some category. For example, we pick values of $i=0,1$ or $2$, which identify the three levels of the categorical factor, and we may sort all data function into three groups based on the value of $i$ used in $y_{i,j,k}.$ On the other hand, if the parameter represent a continuous factor, we pick an interval and the value of the parameter would be randomly drawn from this interval. For example we pick $k\in [0,100]$ and for every data function we use random $k$ from this interval to evaluate $y_{i,j,k}$, the corresponding continuous factor is then equal to $k$.

\begin{figure}%[H]
\includegraphics[scale=0.5]{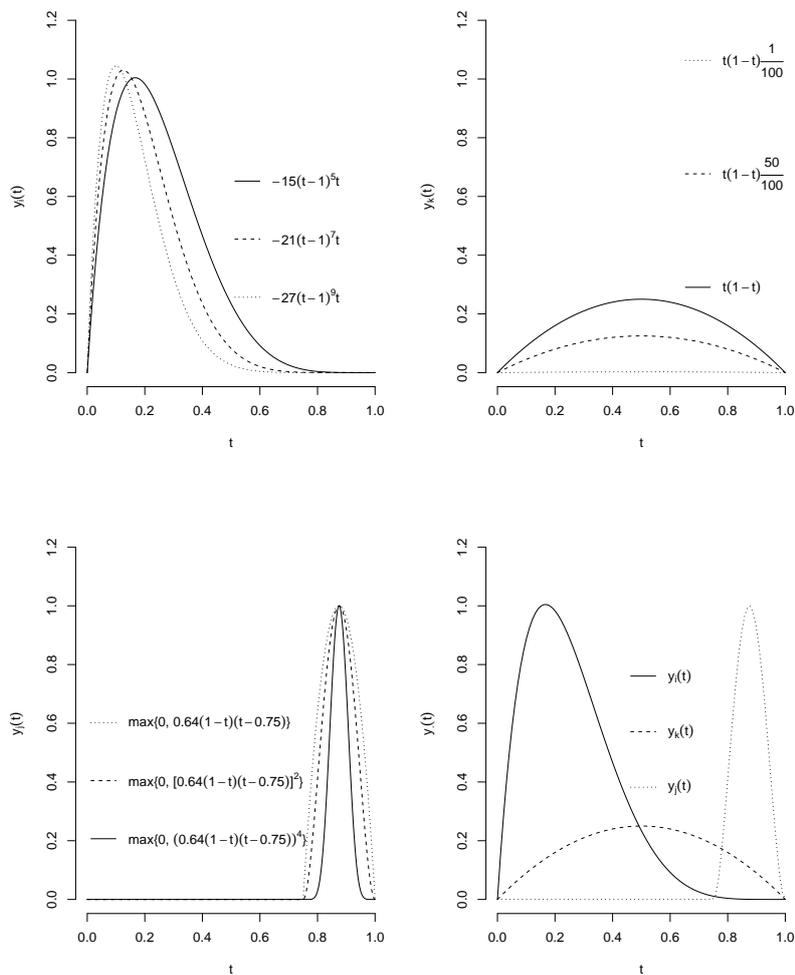}
\caption{Three factors of simulation: In first three quadrants the effects of parameters $i,j$, and $k$ on corresponding factors are visualised; in fourth quadrant the differences between the supports of effects in dependence on $t$ are shown.}\label{figure:Efekty}
\end{figure}

For a better visualisation, we present Figure \ref{figure:Efekty}. In the first three quadrants, we pick a factor and three typical values for the parameter corresponding to the factor in question. The function representing the effect of the parameter is shown. Function $y_i$ based on parameter $i$ effects much more the values for $t<0.5$ and its effect is fading as we enclose $t=1$, function $y_j$ based on the value of $j$ effects only the interval $t\in[0.75,1]$ and $f_k$ based on $k$ effects all values of $t$, but the most significant influence is always in the middle close to $t=0.5$. We would recall these observations later, as we present the sensitivity of the GET to these properties. In the last, fourth, quadrant, we put all the factors in one figure to show the comparison of values and the supports of the functions representing the factors. 

The error term $e(t)$ is chosen to be an i.i.d. error of zero mean and standard deviation equal to $\sigma$. In the Appendix, we also give the results of the simulation study for the Brownian motion error term. Each example consists of 60 simulated data functions, and the significance levels and powers of studied tests are computed from 1000 repetition of the experiment on the preset significance level $\alpha=0.05$. The $p$-value of every permutation test is calculated from 1000 permutations. The RPM method is computed from 30 random projections. Four tests are applied on testing of significance of the factor of interest. The studied tests are our graphical test based on the regression parameters $\mathbf{T}$ (GETP), our graphical test based on based on differences of regression parameters $\mathbf{T}'$ (GETDP), Random projection method (RPM) and $F$-max test.

\subsection{Categorical factor of interest in main effect GLM}
Let us consider two categorical factors, the first one with 3 levels being of interest and the other one with 2 levels being a nuisance factor. This setting produces 6 groups of functions, each one defined by a unique triplet of parameters $i,j,k$ and consisting of 10 functions created by adding an error terms $e(t)$. The estimated probability of rejection is shown in Table \ref{t:1} for 4 tests and 6 different models. In the left column, we show the setting of parameters $i,j,k$ in six triplets generating 6 groups of functions. The factor of interest is amplified by bold font, and we note by ${\bf F1}=0$ that the factor of interest has no effect in the model and by ${\bf F1}\neq 0$ that the factor of interest have some effect in the model. Similarly, we use this notation for nuisance factors $F2$. In the second column, we mark the row with the method which result is presented in the following columns. In the top row, we show the choice of the standard deviation of i.i.d. error. 

\begin{table}[!ht]\caption{The estimated probabilities of rejecting the factor of interest in main effect FGLM with two categorical factors with significance level $\alpha=0.05$.}\label{t:1}
\noindent\begin{tabular}{||l|c||c|c|c||}
\hline
\hline
&Method &$\sigma(e)=0.3$&$\sigma(e)=0.5$&$\sigma(e)=0.8$\\
\hline
\hline
1.{\begin{tabular}{l}
\bf{i}=(1,1,1,1,1,1)\\
j=(1,1,1,1,1,1)\\
k=(1,1,1,1,1,1) \\
${\bf F1}=0,F2=0$
\end{tabular}}
&{\begin{tabular}{l}
GETP\\
GETDP\\
RPM \\
F-max
\end{tabular}}
&{\begin{tabular}{l}
$0.059$\\
$0.071$\\
$0.058$\\
$0.064$
\end{tabular}}
&{\begin{tabular}{l}
$0.057$\\
$0.059$\\
$0.062$\\
$0.055$
\end{tabular}}
&{\begin{tabular}{l}
$0.070$\\
$0.059$\\
$0.073$\\
$0.044$ 
\end{tabular}}
\\ 
\hline
\hline
2.{\begin{tabular}{l}
\bf{i}=(1,1,1,1,1,1)\\
j=(1,1,1,1,1,1)\\
k=(1,1,1,50,50,50) \\
${\bf F1}=0,F2\neq 0$
\end{tabular}}&{\begin{tabular}{l}
GETP\\
GETDP\\
RPM \\
F-max
\end{tabular}}
&{\begin{tabular}{l}
$0.052$\\
$0.069$\\
$0.069$\\
$0.037$
\end{tabular}}
&{\begin{tabular}{l}
$0.050$\\
$0.056$\\
$0.061$\\
$0.041$
\end{tabular}}
&{\begin{tabular}{l}
$0.063$\\
$0.068$\\
$0.087$\\
$0.050$
\end{tabular}}
\\ 
\hline
\hline
3.{\begin{tabular}{l}
\bf{i}=(0,1,2,0,1,2)\\
j=(1,1,1,1,1,1)\\
k=(1,1,1,1,1,1) \\
${\bf F1}\neq 0,F2=0$
\end{tabular}}
&{\begin{tabular}{l}
GETP\\
GETDP\\
RPM \\
F-max
\end{tabular}}
&{\begin{tabular}{l}
$1.000$\\
$1.000$\\
$1.000$\\
$1.000$
\end{tabular}}
&{\begin{tabular}{l}
$0.996$\\
$0.997$\\
$0.766$\\
$0.893$
\end{tabular}}
&{\begin{tabular}{l}
$0.594$\\
$0.602$\\
$0.322$\\
$0.329$
\end{tabular}}
\\ 
\hline
\hline
4.{\begin{tabular}{l}
\bf{i}=(0,1,2,0,1,2)\\
j=(1,1,1,1,1,1)\\
k=(1,1,1,50,50,50) \\
${\bf F1}\neq0,F2\neq0$
\end{tabular}}
&{\begin{tabular}{l}
GETP\\
GETDP\\
RPM \\
F-max
\end{tabular}}
&{\begin{tabular}{l}
$1.000$\\
$1.000$\\
$0.999$\\
$1.000$
\end{tabular}}
&{\begin{tabular}{l}
$0.992$\\
$0.996$\\
$0.778$\\
$0.894$
\end{tabular}}
&{\begin{tabular}{l}
$0.623$\\
$0.609$\\
$0.340$\\
$0.323$
\end{tabular}}
\\ 
\hline
\hline
5.{\begin{tabular}{l}
i=(1,1,1,1,1,1)\\
\bf{j}=(1,2,4,1,2,4)\\
k=(1,1,1,1,1,1)\\
${\bf F1}\neq0,F2=0$
\end{tabular}}
&{\begin{tabular}{l}
GETP\\
GETDP\\
RPM \\
F-max
\end{tabular}}
&{\begin{tabular}{l}
$1.000$\\
$1.000$\\
$0.907$ \\
$1.000$
\end{tabular}}
&{\begin{tabular}{l}
$0.932$\\
$0.940$\\
$0.427$ \\
$0.772$
\end{tabular}}
&{\begin{tabular}{l}
$0.366$\\
$0.370$\\
$0.183$\\
$0.224$
\end{tabular}}
\\ 
\hline
\hline
6.{\begin{tabular}{l}
i=(1,1,1,1,1,1)\\
\bf{j}=(1,2,4,1,2,4)\\
k=(1,1,1,50,50,50)\\
${\bf F1}\neq0,F2\neq0$
\end{tabular}}
&{\begin{tabular}{l}
GETP\\
GETDP\\
RPM \\
F-max
\end{tabular}}
&{\begin{tabular}{l}
$1.000$\\
$1.000$\\
$0.905$\\
$1.000$
\end{tabular}}
&{\begin{tabular}{l}
$0.920$\\
$0.928$\\
$0.389$\\
$0.781$
\end{tabular}}
&{\begin{tabular}{l}
$0.356$\\
$0.374$\\
$0.177$\\
$0.222$
\end{tabular}}
\\ 
\hline
\hline
\end{tabular}
\end{table}

Let us mention that the effects of first and second factor influence the value of $y(t)$ on the same functional domain in case 4. whereas on the different functional domain for the last setting, since supports of functions $y_i$ and $y_j$ do not overlap.

The Table \ref{t:1} shows that the estimated significance levels (case 1. and 2.) are slightly bigger than the desired 0.05 level in tests using Freedman-Lane permutation scheme. On the other hand, the estimated power (cases 3., 4., 5., 6.)  is much higher for our two tests than for the other two.

\begin{figure}
\begin{center}
\includegraphics[scale=0.45]{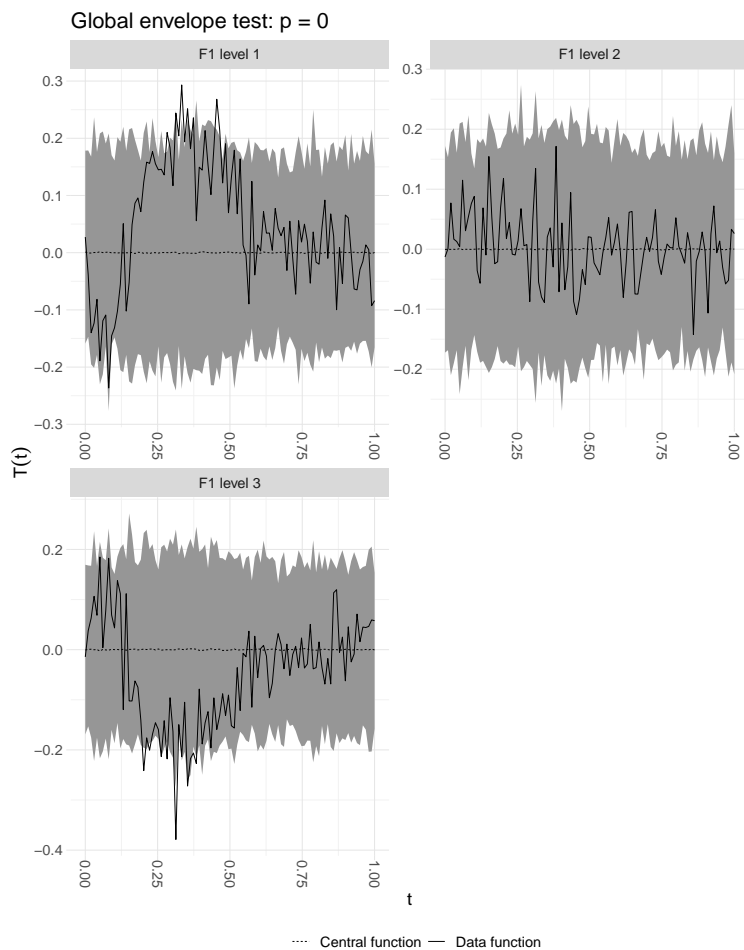}
\end{center}
\caption{GET results for Model 4, $\sigma(e)=0.3$ test vector $\bf{T}$; The first envelope plots the estimated parameters $\beta_{1k}$ based on the choice $i=0$, second envelope plots $i=1$, third envelope plots $i=2$. The values where data function exits the grey area of the 95\% global envelope are responsible for the rejection of null hypothesis. }\label{figure2}
\end{figure}

\begin{figure}
\begin{center}
\includegraphics[scale=0.45]{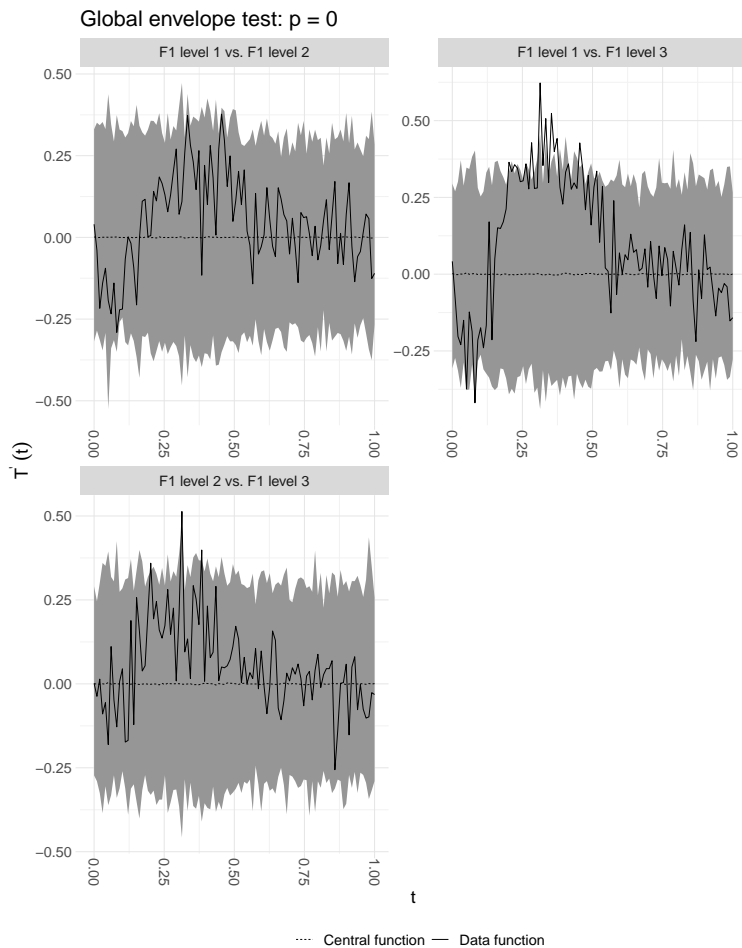}
\end{center}
\caption{GET results for Model 4, $\sigma(e)=0.3$ test vector $\bf{T}'$; The first envelope plots the differences of the estimated parameters $\beta_{1k}$ and $\beta_{2k}$, second envelope plots the difference between the $\beta_{1k}$ and $\beta_{3k}$, third envelope plots the difference between $\beta_{2k}$ and $\beta_{3k}$. The values where data function exits the grey area of the 95\% global envelope are responsible for the rejection of null hypothesis.}\label{figure3}
\end{figure}

We also show the graphical output of our tests for one realisation of model 4 from Table \ref{t:1}, for $\sigma(e)=0.3$. The global 95\% envelope is drawn in grey, and the test vector is drawn in black. We recall the definitions of test vectors from Subsection \ref{DefinujemeTestoveVektory}, first we present graphical outcome of test vectors $\bf{T}$ in Figure \ref{figure2} for three different groups determined by choice of parameter of interest $i=0$, $i=1$ or $i=2$. Figure \ref{figure3} pictures test vector $\mathbf{T}'$ for comparison between three groups of data function determined by $i=0, i=1$ or $i=2$. We should mention, that all figures are plotted for 5000 permutations instead of 1000 permutations result presented in Tables, the reason is that envelope is more smooth for more permutation used.  

The both presented tests reject the null hypothesis with $p$-value < 0.001. In addition the GET method gives us information about the values of $t$ and groups responsible for rejection. As we observe in Figure \ref{figure2}, there are values $t\in (0.2,0.5)$ in the group $i=0$ and group $i=2$ responsible for rejection. We may expect based on figure, that for values $i=0$ the values of functions $y_i$ is bigger than average in responsible interval $t\in (0.2,0.5)$ and for functions in group $i=2$ we expect $y_i$ to be lower than average for $t\in (0.2,0.5)$. As we compare this observation to Figure \ref{figure:Efekty}, we see that we catch the main difference between the shape of $y_i$ for $i=0, i=1$ or $i=2$, that is the shift of the maximum of the function that cause the difference between the functions approximately in interval $t\in(0.2,0.5)$. The similar outcome can be read from Figure \ref{figure3}, where most significant differences are between the groups corresponding to choices $i=0$ and $i=2$, and we see that the difference between them is obvious in interval $(0.2, 0.5)$ and also possibly around $t=0.1$. That is exactly where the distance between the values of $y_i$ for choices $i=0$ and $i=2$ are the biggest. We may also observe some differences between the pairs of groups $i=0, i=1$ or $i=1, i=2$, but the difference there is not that obvious. The big variability of the tested functions is caused by the nature of i.i.d. error term.

\subsection{Continuous factor of interest in main effect GLM}
Let us consider now the continuous factor as the factor of interest. We may consider a nuisance factor to be a categorical factor or another continuous factor. 
The estimated probability of rejection is shown in Table \ref{t:2} for 4 tests and 6 different models in a similar way as above.
 In models 1., 2. the tested continuous factor should not influence the functions, so the parameter $k$ is kept constant. In models 3., 4., 5. and 6. we draw $k$ randomly from the interval $k\in[0,100]$ for each function separately in order to achieve a continuous effect. In model 5. we have a continuous nuisance factor, so we also draw $i\in[0,2]$ randomly. Model 4. has a nuisance effect in the same domain as the effect of interest, model 6. is analogous, but the nuisance effect is in a different domain than the factor of interest.
 
 The GETDP method is not applicable in this case since we have only one parameter for the continuous effect. The R implementation of RPM is not designed for the continuous factors. Therefore Table \ref{t:2} shows results only for the other two methods. Similarly like in the previous study the estimated significant level could be a bit higher than the nominal level for the GETP, but the power is much higher for the GETP than for the $F$-max procedure.

\begin{table}[!ht]\caption{The estimated probabilities of rejecting of factor of interest in main effect two factor FGLM with continuous factor of interest.} \label{t:2}
\noindent\begin{tabular}{||l|c||c|c|c||c|c|c|c||}
\hline
\hline
&Method&$\sigma(e)=0.3$&$\sigma(e)=0.5$&$\sigma(e)=0.8$\\
\hline
\hline
1.{\begin{tabular}{l}
i=(1,1,1,1,1,1)\\
j=(1,1,1,1,1,1)\\
\bf{k}=(1,1,1,1,1,1)\\
${\bf F1}=0,F2=0$
\end{tabular}}
&{\begin{tabular}{l}
GETP\\
GETDP\\
RPM \\
F-max
\end{tabular}}
&{\begin{tabular}{c}
$0.045$\\
$-$\\
$-$\\
$0.049$
\end{tabular}}
&{\begin{tabular}{c}
$0.057$\\
$-$\\
$-$\\
$0.058$
\end{tabular}}
&{\begin{tabular}{c}
$0.056$\\
$-$\\
$-$\\
$0.059$
\end{tabular}}
\\
\hline
\hline
2.{\begin{tabular}{l}
i=(0,1,2,0,1,2)\\
j=(1,1,1,1,1,1)\\
\bf{k}=(1,1,1,1,1,1)\\
${\bf F1}=0,F2\neq 0$
\end{tabular}}
&{\begin{tabular}{l}
GETP\\
GETDP\\
RPM \\
F-max
\end{tabular}}
&{\begin{tabular}{c}
$0.028$\\
$-$\\
$-$\\
$0.047$
\end{tabular}}
&{\begin{tabular}{c}
$0.076$\\
$-$\\
$-$\\
$0.064$
\end{tabular}}
&{\begin{tabular}{c}
$0.072$\\
$-$\\
$-$ \\
$0.042$
\end{tabular}}
\\
\hline
\hline
3.{\begin{tabular}{c}
i=(1,1,1,1,1,1)\\
j=(1,1,1,1,1,1)\\
{\bf k$\in$[0,100]} \\
${\bf F1}\neq 0,F2=0$
\end{tabular}}&{\begin{tabular}{l}
GETP\\
GETDP\\
RPM \\
F-max
\end{tabular}}
&{\begin{tabular}{c}
$0.970$\\
$-$\\
$-$ \\
$0.800$
\end{tabular}}
&{\begin{tabular}{c}
$0.626$\\
$-$\\
$-$\\
$0.384$
\end{tabular}}
&{\begin{tabular}{c}
$0.202$\\
$-$\\
$-$\\
$0.152$
\end{tabular}}
\\
\hline
\hline
4.{\begin{tabular}{c}
i=(0,1,2,0,1,2)\\
j=(1,1,1,1,1,1)\\
{\bf k$\in$[0,100]} \\
${\bf F1}\neq0,F2\neq0$
\end{tabular}}
&{\begin{tabular}{l}
GETP\\
GETDP\\
RPM \\
F-max
\end{tabular}}
&{\begin{tabular}{c}
$0.704$\\
$-$\\
$-$ \\
$0.435$
\end{tabular}}
&{\begin{tabular}{c}
$0.355$\\
$-$\\
$-$ \\
$0.233$
\end{tabular}}
&{\begin{tabular}{c}
$0.208$\\
$-$\\
$-$\\
$0.126$
\end{tabular}}
\\
\hline
\hline
5.{\begin{tabular}{c}
i$\in$[0,2]\\
j=(1,1,1,1,1,1)\\
{\bf k$\in$[0,100]}\\
${\bf F1}\neq0,F2=0$
\end{tabular}}
&{\begin{tabular}{l}
GETP\\
GETDP\\
RPM \\
F-max
\end{tabular}}
&{\begin{tabular}{c}
$0.981$\\
$-$\\
$-$ \\
$0.857$
\end{tabular}}
&{\begin{tabular}{c}
$0.493$\\
$-$\\
$-$\\
$0.262$
\end{tabular}}
&{\begin{tabular}{c}
$0.199$\\
$-$\\
$-$ \\
$0.132$
\end{tabular}}
\\
\hline
\hline
6.{\begin{tabular}{c}
i=(1,1,1,1,1,1)\\
j=(1,2,4,1,2,4)\\
{\bf k$\in$[0,100]}\\
${\bf F1}\neq0,F2\neq0$
\end{tabular}}
&{\begin{tabular}{l}
GETP\\
GETDP\\
RPM \\
F-max
\end{tabular}}
&{\begin{tabular}{c}
$0.986$\\
$-$\\
$-$\\
$0.841$
\end{tabular}}
&{\begin{tabular}{c}
$0.53$\\
$-$\\
$-$\\
$0.325$
\end{tabular}}
&{\begin{tabular}{c}
$0.237$\\
$-$\\
$-$\\
$0.175$
\end{tabular}}
\\
\hline
\hline
\end{tabular}
\end{table}

\begin{figure}
\includegraphics[scale=0.45]{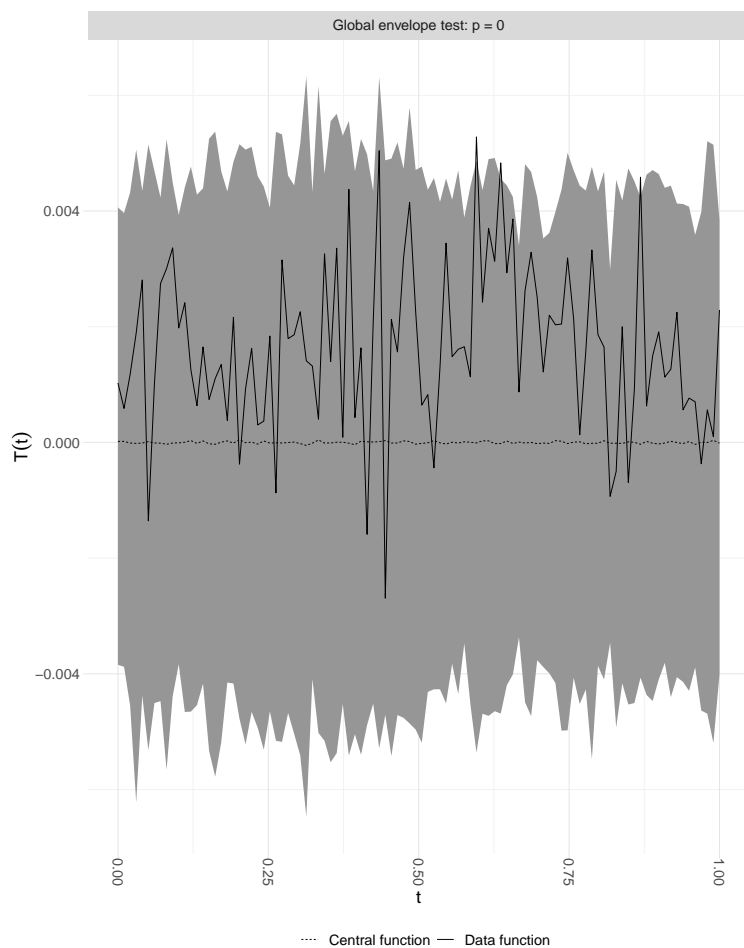}
\caption{GET results for one realisation of model 4 from Table \ref{t:2}, $\sigma(e)=0.3$. The values where data function exits the grey area of the 95\% global envelope are responsible for the rejection of null hypothesis. }\label{FigCOnt}
\end{figure}

Once again we demonstrate the visual outcome of the GET method by plotting the envelope in Figure \ref{FigCOnt} for one realisation of model 4 from Table \ref{t:2}. We do not search for a significant group as previous, but we try to find the values of $t$ responsible for rejection, it means the values where the continuous effect has the most influence on the functions. In Figure \ref{FigCOnt}, these are the ones that exit the envelope, and we see that it is the values in the middle of the interval, $t=0.5$ and values close to it. We compare this result of envelope test to the model function plot in Figure \ref{figure:Efekty}, and clearly, we observe that $y_k$ has the most extreme values in the middle of the interval, as we expect based on visualisation given by envelope test.

\subsection{Interactions}
Let us consider now the factorial model of two categorical factors, the first one with 3 levels and the other one with 2 levels. The factor of interest are now interactions between these two factors. We get 6 groups of functions again, each of them consisting of 10 functions and defined by a unique triplet of parameters $i,j,k$ and a random error term. The estimated probability of rejection is shown in Table \ref{t:3} for 4 tests and 4 different models. The factor of interest is amplified by bold font, and we note by ${\bf I}=0$ that the interactions are not present in the model and by ${\bf I}\neq 0$ that the interactions are present in the model.

Here we observe the slight liberality of the methods GETP, GETDP and RPM and again much higher estimated powers for our methods than the other two methods.

\begin{table}[!ht]\caption{The estimated probabilities of rejecting of effect of interactions in factorial two factor FGLM.}\label{t:3}
\noindent\begin{tabular}{||l|c||c|c|c||}
\hline
\hline
&Method&$\sigma(e)=0.3$&$\sigma(e)=0.5$&$\sigma(e)=0.8$\\
\hline
\hline
1.{\begin{tabular}{l}
\bf{i}=(0,1,2,0,1,2)\\
j=(1,1,1,1,1,1)\\
k=(1,1,1,50,50,50) \\
${\bf I}=0$
\end{tabular}}
&{\begin{tabular}{l}
GETP\\
GETDP\\
RPM \\
F-max
\end{tabular}}
&{\begin{tabular}{l}
$0.084$\\
$0.084$\\
$0.076$\\
$0.060$
\end{tabular}}
&{\begin{tabular}{l}
$0.077$\\
$0.078$\\
$0.069$\\
$0.049$
\end{tabular}}
&{\begin{tabular}{l}
$0.095$\\
$0.084$\\
$0.078$\\
$0.052$
\end{tabular}}\\
\hline
\hline
2.{\begin{tabular}{l}
\bf{i}=(0,1,2,1,1,1)\\
j=(1,1,1,1,1,1)\\
k=(1,1,1,50,50,50) \\
${\bf I}\neq 0$
\end{tabular}}
&{\begin{tabular}{l}
GETP\\
GETDP\\
RPM \\
F-max
\end{tabular}}
&{\begin{tabular}{l}
$0.961$\\
$0.962$\\
$0.588$\\
$0.652$
\end{tabular}}
&{\begin{tabular}{l}
$0.457$\\
$0.440$\\
$0.222$\\
$0.192$
\end{tabular}}
&{\begin{tabular}{l}
$0.184$\\
$0.182$\\
$0.118$\\
$0.096$
\end{tabular}}\\
\hline
\hline
3.{\begin{tabular}{l}
\bf{i}=(0,1,2,0,1,2)\\
j=(1,1,1,2,2,2)\\
k=(1,1,1,1,1,1) \\
${\bf I}= 0$
\end{tabular}}
&{\begin{tabular}{l}
GETP\\
GETDP\\
RPM \\
F-max
\end{tabular}}
&{\begin{tabular}{l}
$0.093$\\
$0.087$\\
$0.079$\\
$0.049$
\end{tabular}}
&{\begin{tabular}{l}
$0.078$\\
$0.081$\\
$0.070$\\
$0.052$
\end{tabular}}
&{\begin{tabular}{l}
$0.075$\\
$0.070$\\
$0.070$\\
$0.043$
\end{tabular}}\\
\hline
\hline
4.{\begin{tabular}{l}
\bf{i}=(0,1,2,1,1,1)\\
j=(1,1,1,2,2,2)\\
k=(1,1,1,1,1,1) \\
${\bf I}\neq0$
\end{tabular}}
&{\begin{tabular}{l}
GETP\\
GETDP\\
RPM \\
F-max
\end{tabular}}
&{\begin{tabular}{l}
$0.959$\\
$0.951$\\
$0.597$\\
$0.647$
\end{tabular}}
&{\begin{tabular}{l}
$0.425$\\
$0.425$\\
$0.218$\\
$0.199$
\end{tabular}}
&{\begin{tabular}{l}
$0.188$\\
$0.189$\\
$0.110$\\
$0.086$ 
\end{tabular}}\\
\hline
\hline
\end{tabular}
\end{table}

\begin{figure}
\includegraphics[scale=0.45]{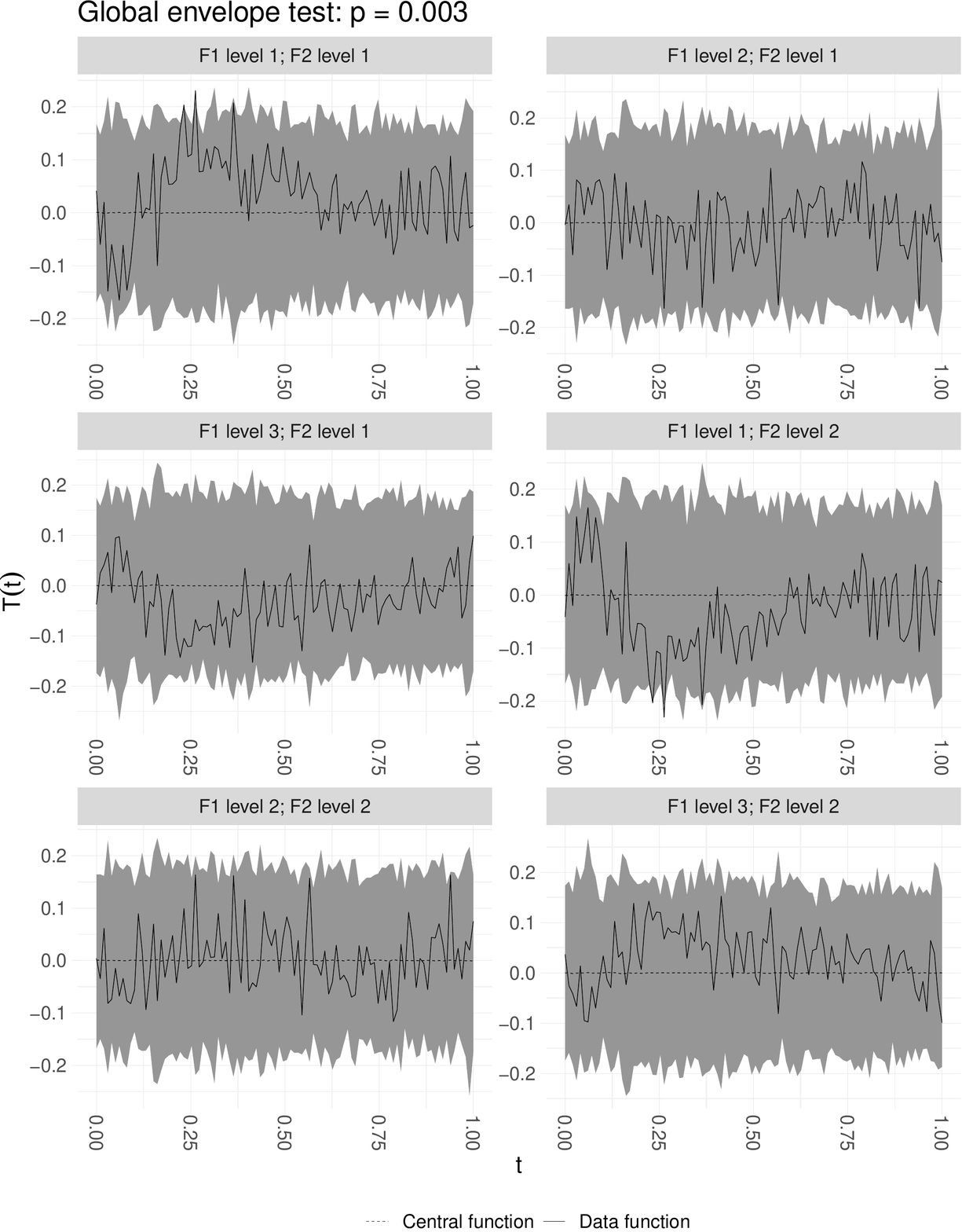}
\caption{GET results for one realisation of model 4 from Table \ref{t:3}, $\sigma(e)=0.3$, with test vector $\bf{T}$. }\label{f:5}
\end{figure}

Figure \ref{f:5} shows the graphical results of the GETP test for one realisation of the model 4 from Table \ref{t:3}. The estimated parameter functions leave the envelope for factor 1 level 1 groups only. This outcome identifies the differences between the groups of factor 2 level1 and factor 2 level 2 in the level 1 of factor 1. Looking in Table \ref{t:3} we should see also the differences in factor 1 level 3 but since the differences between functions with $i=1$ and $i=2$ are smaller than the differences between functions with $i=0$ and $i=1$ we do not identify this difference.

\begin{figure}
\includegraphics[scale=0.65]{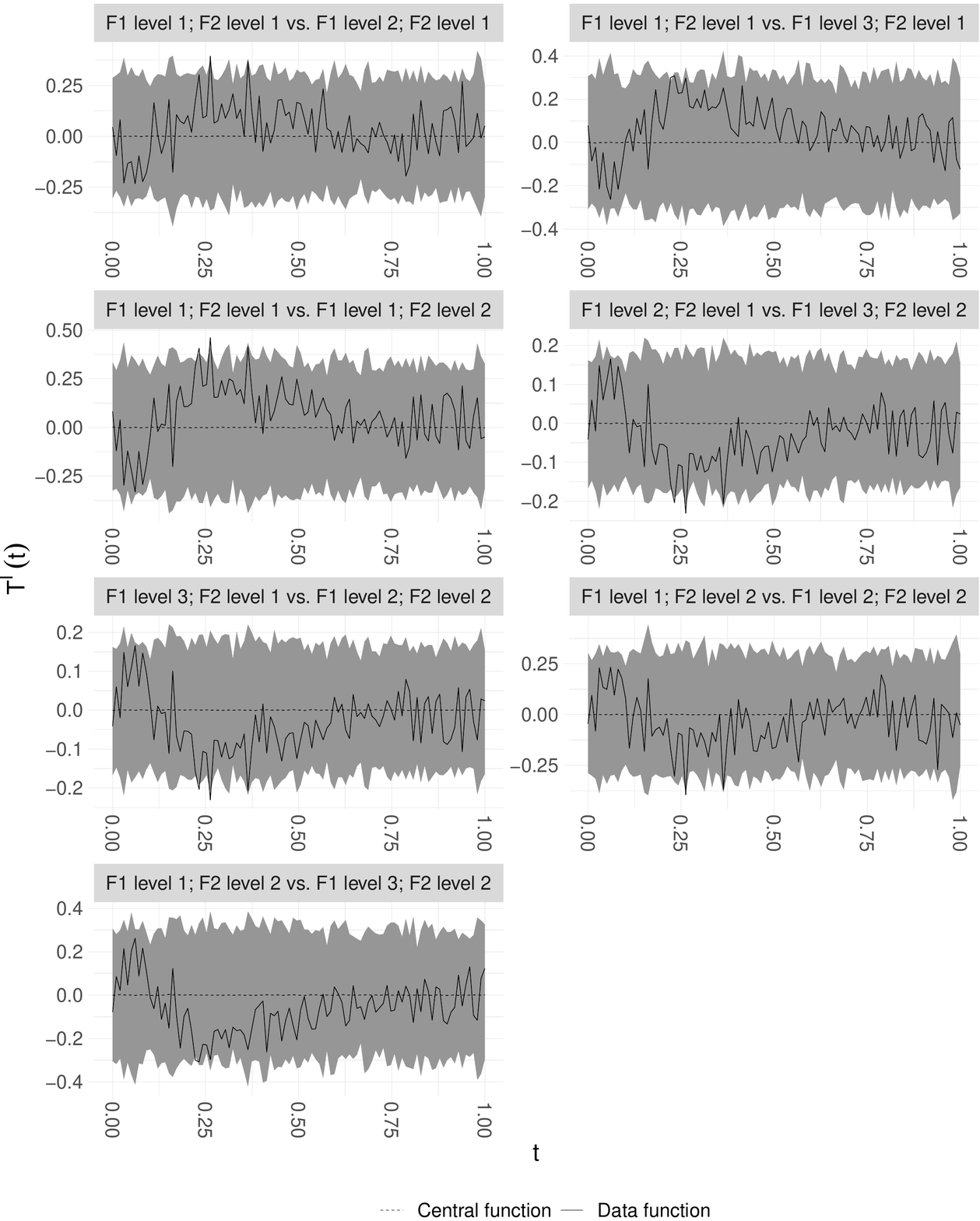}
\caption{GET results for one realisation of model 4 from Table \ref{t:3}, $\sigma(e)=0.3$, with test vector $\bf{T'}$. Due to the high number of comparisons, only the significant are presented.} \label{f:6}
\end{figure}

Figure \ref{f:6} shows the graphical results of the GETDP test for the same realisation as above. Here we identify the differences between groups of functions identified by factor 1 level 1 and several other groups, as in the previous image. Furthermore, there is identified the difference between factor 1 level 3 groups and a few other groups, which was not identified by GETP test.  

\section{Discussion and conclusions}\label{sec:DC}
In this paper, we introduced a nonparametric test of significance in the functional general linear model with either categorical and continuous factors and also with interactions. The simulation studies performed in cases of interactions, categorical and continuous factors of interest show that our proposed tests are very slightly liberal, due to the application of Freedman-Lane permutation scheme in the presence of nuisance factors. On the other hand, they show much bigger power than random projection method and the $F$-max method. The differences between powers of the tests can be even bigger when the random error is not i.i.d., see Appendix.

The advantage of our method in comparison to other methods is the graphical interpretation which it provides, especially it is able to detect the functional domain which is responsible for the potential rejection. The graphical interpretation is the primary purpose why we propose this method. Also, its base in rank statistics makes it robust with respect to changes of the distribution of the test statistics across the domain of the functions. E.g. $F$ statistic does not change its first and second moments across the domain but if the original functional data are not Gaussian than the other moments and especially the quantiles of the $F$ statistic changes across the domain. In such cases, the $F$-max statistic can be blind to deviation from the null model in some parts of the domain. The size of the robustness of rank based method concerning $F$-max method will be studied in the future.

Finally, the post hoc nature of our test based on the differences of regression parameters is another advantage by providing the better interpretation capabilities of our method. %Besides the simulation results suggest promoted method may produce significantly improved results in comparison to classical methods, which is surprising since the primary goal of the method is visualisation and finding responsible values of $t$ and responsible category.

%\bibliographystyle{chicago}
%\bibliography{sample}
\begin{acknowledgements}
The project has been financially supported by the Grant Agency of Czech Republic (Project No. 19-04412S).
\end{acknowledgements}
% etc
%\end{thebibliography}

\bibliographystyle{spbasic}      % basic style, author-year citations
\bibliography{sample}   % name your BibTeX data base

%\begin{thebibliography}{}

\section*{Appendix}
Let us consider the previous simulation design, where the i.i.d. error term  $e(t)$ would be replaced by the Brownian motion. The difference is that with i.i.d. error used in previous sections the variance is constant, but with the Brownian motion, it is increasing in dependence on $t$. This may cause some trouble, since the bigger variance for bigger $t$ means different sensitivity for effects influencing values close to $t=0$, such as parameter $i$ and effects that influence the values close to $t=1$ such as parameter $j$, see Figure \ref{figure:Efekty}.

%\begin{verbatim}
%# funkce generujici chybovou slozku
%error <- function(Y, type = 'normal', sigma = 0.3) {
%  n <- dim(Y)[1]
%  p <- dim(Y)[2]
%  if (type == 'normal') {
%    error <- replicate(p, rnorm(n, 0, sigma))
%  } else {
%    if (type == 'walk') {
%      error <- replicate(p, cumsum(rnorm(n, 0, sigma)))
%    }
%    mu <- rep(0, p)
%    sigmas <- rep(sigma, p) # pripadne tady dat sigmas jako vektor sigma 
%                    #jehoz elementy umoznuji nastavit ruzne rozptyly pro %jednotlive t
%    error <- mapply(function(x, y) {
%      rnorm(x, y, n = n)
%    }, x = mu, y = sigmas)
%  }
%  error
%}    
    
%\end{verbatim}

The standard deviation of the Brownian motion $e(1)$ was kept 10 times bigger than the standard deviation of the i.i.d. error, since then the increments in our discrete Brownian motion has the same standard deviation as the i.i.d. error and we get comparable results. 

%{In the first row of tables, where we have $\sigma(e)$ in previous simulations, we present a deviation in one step.} But there are 100 steps, as the total variance of error is adding over all previous steps for all previous values of $t$. We still keep the notation $\sigma(e)$, but in fact, it is the deviation of the difference between the errors in the two following steps. We do not calculate the exact values of this deviation in dependence on $t$ since the result is too technical and it is not the aim of this research.

{We present 3 tables in the same spirit as in the main text. The results here are calculated from $100$ simulations only since we did not have enough time to finish the whole study. The full study will appear in the final version.}

The estimated levels of significance are slightly liberal for the procedures using the Freedman-Lane algorithm. The powers of our tests are again much bigger than the powers of the other two tests. Even more, in some cases, the difference between these tests is more significant than for the i.i.d. error rate and in other cases the difference between these tests is similar as for the i.i.d. error rate.

\begin{table}[!ht]\caption{The estimated probabilities of rejecting of factor of interest in main effect FGLM with two categorical factors and Brownian motion error.}\label{t:4}
\noindent\begin{tabular}{||l|c||c|c|c||}
\hline
\hline
&Method&$\sigma(e(1))=3$&$\sigma(e(1))=5$&$\sigma(e(1))=8$\\
\hline
\hline
1.{\begin{tabular}{l}
\bf{i}=(1,1,1,1,1,1)\\
j=(1,1,1,1,1,1)\\
k=(1,1,1,1,1,1) \\
${\bf F1}=0,F2=0$
\end{tabular}}
&{\begin{tabular}{l}
GETP\\
GETDP\\
RPM \\
F-max
\end{tabular}}
&{\begin{tabular}{l}
$0.03$\\
$0.06$\\
$0.03$\\
$0.03$
\end{tabular}}
&{\begin{tabular}{l}
$0.08$\\
$0.06$\\
$0.09$\\
$0.06$
\end{tabular}}
&{\begin{tabular}{l}
$0.05$\\
$0.02$\\
$0.07$\\
$0.02$
\end{tabular}}
\\ 
\hline
\hline
2.{\begin{tabular}{l}
\bf{i}=(1,1,1,1,1,1)\\
j=(1,1,1,1,1,1)\\
k=(1,1,1,50,50,50) \\
${\bf F1}=0,F2\neq 0$
\end{tabular}}&{\begin{tabular}{l}
GETP\\
GETDP\\
RPM \\
F-max
\end{tabular}}
&{\begin{tabular}{l}
$0.08$\\
$0.06$\\
$0.06$\\
$0.07$
\end{tabular}}
&{\begin{tabular}{l}
$0.08$\\
$0.09$\\
$0.07$\\
$0.09$
\end{tabular}}
&{\begin{tabular}{l}
$0.03$\\
$0.00$\\
$0.08$\\
$0.06$
\end{tabular}}
\\ 
\hline
\hline
3.{\begin{tabular}{l}
\bf{i}=(0,1,2,0,1,2)\\
j=(1,1,1,1,1,1)\\
k=(1,1,1,1,1,1) \\
${\bf F1}\neq 0,F2=0$
\end{tabular}}
&{\begin{tabular}{l}
GETP\\
GETDP\\
RPM \\
F-max
\end{tabular}}
&{\begin{tabular}{l}
$1.00$\\
$1.00$\\
$1.00$ \\
$1.00$
\end{tabular}}
&{\begin{tabular}{l}
$1.00$\\
$1.00$\\
$0.77$ \\
$0.90$
\end{tabular}}
&{\begin{tabular}{l}
$0.76$\\
$0.73$\\
$0.33$ \\
$0.23$
\end{tabular}}
\\ 
\hline
\hline
4.{\begin{tabular}{l}
\bf{i}=(0,1,2,0,1,2)\\
j=(1,1,1,1,1,1)\\
k=(1,1,1,50,50,50) \\
${\bf F1}\neq0,F2\neq0$
\end{tabular}}
&{\begin{tabular}{l}
GETP\\
GETDP\\
RPM \\
F-max
\end{tabular}}
&{\begin{tabular}{l}
$1.00$\\
$1.00$\\
$1.00$ \\
$1.00$
\end{tabular}}
&{\begin{tabular}{l}
$1.00$\\
$1.00$\\
$0.82$ \\
$0.83$
\end{tabular}}
&{\begin{tabular}{l}
$0.71$\\
$0.67$\\
$0.28$ \\
$0.38$
\end{tabular}}
\\ 
\hline
\hline
5.{\begin{tabular}{l}
i=(1,1,1,1,1,1)\\
\bf{j}=(1,2,4,1,2,4)\\
k=(1,1,1,1,1,1)\\
${\bf F1}\neq0,F2=0$
\end{tabular}}
&{\begin{tabular}{l}
GETP\\
GETDP\\
RPM \\
F-max
\end{tabular}}
&{\begin{tabular}{l}
$1.00$\\
$1.00$\\
$0.89$ \\
$1.00$
\end{tabular}}
&{\begin{tabular}{l}
$0.90$\\
$0.88$\\
$0.46$ \\
$0.75$
\end{tabular}}
&{\begin{tabular}{l}
$0.34$\\
$0.29$\\
$0.14$ \\
$0.22$
\end{tabular}}
\\ 
\hline
\hline
6.{\begin{tabular}{l}
i=(1,1,1,1,1,1)\\
\bf{j}=(1,2,4,1,2,4)\\
k=(1,1,1,50,50,50)\\
${\bf F1}\neq0,F2\neq0$
\end{tabular}}
&{\begin{tabular}{l}
GETP\\
GETDP\\
RPM \\
F-max
\end{tabular}}
&{\begin{tabular}{l}
$1.00$\\
$1.00$\\
$0.92$ \\
$1.00$
\end{tabular}}
&{\begin{tabular}{l}
$0.87$\\
$0.85$\\
$0.34$ \\
$0.79$
\end{tabular}}
&{\begin{tabular}{l}
$0.34$\\
$0.35$\\
$0.18$\\
$0.25$
\end{tabular}}
\\ 
\hline
\hline
\end{tabular}
\end{table}

\begin{table}[!ht]\caption{The estimated probabilities of rejecting of factor of interest in main effect two factor FGLM with continuous factor of interest and Brownian motion error.} \label{t:5}
\noindent\begin{tabular}{||l|c||c|c|c||c|c|c|c||}
\hline
\hline
&Method&$\sigma(e(1))=3$&$\sigma(e(1))=5$&$\sigma(e(1))=8$\\
\hline
\hline
1.{\begin{tabular}{l}
i=(1,1,1,1,1,1)\\
j=(1,1,1,1,1,1)\\
\bf{k}=(1,1,1,1,1,1)\\
${\bf F1}=0,F2=0$
\end{tabular}}
&{\begin{tabular}{l}
GETP\\
GETDP\\
RPM \\
F-max
\end{tabular}}
&{\begin{tabular}{l}
$0.04$\\
$-$\\
$-$\\
$0.07$
\end{tabular}}
&{\begin{tabular}{l}
$0.07$\\
$-$\\
$-$ \\
$0.08$
\end{tabular}}
&{\begin{tabular}{l}
$0.07$\\
$-$\\
$-$ \\
$0.02$
\end{tabular}}
\\
\hline
\hline
2.{\begin{tabular}{l}
i=(0,1,2,0,1,2)\\
j=(1,1,1,1,1,1)\\
\bf{k}=(1,1,1,1,1,1)\\
${\bf F1}=0,F2\neq 0$
\end{tabular}}
&{\begin{tabular}{l}
GETP\\
GETDP\\
RPM \\
F-max
\end{tabular}}
&{\begin{tabular}{l}
$0.05$\\
$-$\\
$-$ \\
$0.01$
\end{tabular}}
&{\begin{tabular}{l}
$0.07$\\
$-$\\
$-$ \\
$0.05$
\end{tabular}}
&{\begin{tabular}{l}
$0.046$\\
$-$\\
$-$ \\
$0.040$
\end{tabular}}
\\
\hline
\hline
3.{\begin{tabular}{l}
i=(1,1,1,1,1,1)\\
j=(1,1,1,1,1,1)\\
{\bf k$\in$[0,100]} \\
${\bf F1}\neq 0,F2=0$
\end{tabular}}&{\begin{tabular}{l}
GETP\\
GETDP\\
RPM \\
F-max
\end{tabular}}
&{\begin{tabular}{l}
$0.993$\\
$-$\\
$-$\\
$0.904$
\end{tabular}}
&{\begin{tabular}{l}
$0.495$\\
$-$\\
$-$\\
$0.295$
\end{tabular}}
&{\begin{tabular}{l}
$0.269$\\
$-$\\
$-$\\
$0.168$
\end{tabular}}
\\
\hline
\hline
4.{\begin{tabular}{l}
i=(0,1,2,0,1,2)\\
j=(1,1,1,1,1,1)\\
{\bf k$\in$[0,100]} \\
${\bf F1}\neq0,F2\neq0$
\end{tabular}}
&{\begin{tabular}{l}
GETP\\
GETDP\\
RPM \\
F-max
\end{tabular}}
&{\begin{tabular}{l}
$0.905$\\
$-$\\
$-$\\
$0.667$
\end{tabular}}
&{\begin{tabular}{l}
$0.314$\\
$-$\\
$-$\\
$0.174$
\end{tabular}}
&{\begin{tabular}{l}
$0.123$\\
$-$\\
$-$\\
$0.091$
\end{tabular}}
\\
\hline
\hline
5.{\begin{tabular}{l}
i$\in$[0,2]\\
j=(1,1,1,1,1,1)\\
{\bf k$\in$[0,100]}\\
${\bf F1}\neq0,F2=0$
\end{tabular}}
&{\begin{tabular}{l}
GETP\\
GETDP\\
RPM \\
F-max
\end{tabular}}
&{\begin{tabular}{l}
$0.991$\\
$-$ \\
$-$ \\
$0.874$
\end{tabular}}
&{\begin{tabular}{l}
$0.497$\\
$-$\\
$-$\\
$0.266$
\end{tabular}}
&{\begin{tabular}{l}
$0.206$\\
$-$\\
$-$\\
$0.139$
\end{tabular}}
\\
\hline
\hline
6.{\begin{tabular}{l}
i=(1,1,1,1,1,1)\\
j=(1,2,4,1,2,4)\\
{\bf k$\in$[0,100]}\\
${\bf F1}\neq0,F2\neq0$
\end{tabular}}
&{\begin{tabular}{l}
GETP\\
GETDP\\
RPM \\
F-max
\end{tabular}}
&{\begin{tabular}{l}
$0.955$\\
$-$\\
$-$\\
$0.747$
\end{tabular}}
&{\begin{tabular}{l}
$0.426$\\
$-$\\
$-$\\
$0.264$
\end{tabular}}
&{\begin{tabular}{l}
$0.179$\\
$-$\\
$-$\\
$0.127$
\end{tabular}}
\\
\hline
\hline
\end{tabular}
\end{table}

\begin{table}[!ht]\caption{The estimated probabilities of rejecting of effect of interactions in factorial two factor FGLM with Brownian motion error.}\label{t:6}
\noindent\begin{tabular}{||l|c||c|c|c||}
\hline
\hline
&Method&$\sigma(e(1))=3$&$\sigma(e(1))=5$&$\sigma(e(1))=8$\\
\hline
\hline
1.{\begin{tabular}{l}
\bf{i}=(0,1,2,0,1,2)\\
j=(1,1,1,1,1,1)\\
k=(1,1,1,50,50,50) \\
${\bf I}=0$
\end{tabular}}
&{\begin{tabular}{l}
GETP\\
GETDP\\
RPM \\
F-max
\end{tabular}}
&{\begin{tabular}{l}
$0.09$\\
$0.11$\\
$0.12$\\
$0.05$
\end{tabular}}
&{\begin{tabular}{l}
$0.09$\\
$0.11$\\
$0.07$\\
$0.05$
\end{tabular}}
&{\begin{tabular}{l}
$0.15$\\
$0.09$\\
$0.06$\\
$0.04$
\end{tabular}}\\
\hline
\hline
2.{\begin{tabular}{l}
\bf{i}=(0,1,2,1,1,1)\\
j=(1,1,1,1,1,1)\\
k=(1,1,1,50,50,50) \\
${\bf I}\neq 0$
\end{tabular}}
&{\begin{tabular}{l}
GETP\\
GETDP \\
RPM \\
F-max
\end{tabular}}
&{\begin{tabular}{l}
$0.97$\\
$0.95$\\
$0.62$\\
$0.66$
\end{tabular}}
&{\begin{tabular}{l}
$0.45$\\
$0.46$\\
$0.19$\\
$0.24$
\end{tabular}}
&{\begin{tabular}{l}
$0.22$\\
$0.17$\\
$0.13$\\
$0.13$
\end{tabular}}\\
\hline
\hline
3.{\begin{tabular}{l}
\bf{i}=(0,1,2,0,1,2)\\
j=(1,1,1,2,2,2)\\
k=(1,1,1,1,1,1) \\
${\bf I}=0$
\end{tabular}}
&{\begin{tabular}{l}
GETP\\
GETDP\\
RPM \\
F-max
\end{tabular}}
&{\begin{tabular}{l}
$0.10$\\
$0.10$\\
$0.08$\\
$0.02$
\end{tabular}}
&{\begin{tabular}{l}
$0.06$\\
$0.07$\\
$0.10$\\
$0.06$
\end{tabular}}
&{\begin{tabular}{l}
$0.05$\\
$0.05$\\
$0.05$\\
$0.08$
\end{tabular}}\\
\hline
\hline
4.{\begin{tabular}{l}
\bf{i}=(0,1,2,1,1,1)\\
j=(1,1,1,2,2,2)\\
k=(1,1,1,1,1,1) \\
${\bf I}\neq0$
\end{tabular}}
&{\begin{tabular}{l}
GETP\\
GETDP\\
RPM \\
F-max
\end{tabular}}
&{\begin{tabular}{l}
$1.00$\\
$1.00$\\
$0.55$\\
$0.64$
\end{tabular}}
&{\begin{tabular}{l}
$0.54$\\
$0.58$\\
$0.22$\\
$0.24$
\end{tabular}}
&{\begin{tabular}{l}
$0.28$\\
$0.27$\\
$0.07$ \\
$0.07$
\end{tabular}}\\
\hline
\hline
\end{tabular}
\end{table}

\end{document}